# Proofreading mechanism of Class 2 CRISPR-Cas systems

Aset Khakimzhan[1,*], David Garenne[1], Benjamin I. Tickman[2], Jason Fontana[2], James Carothers[2,3,4], Vincent Noireaux[1,*]

[1]School of Physics and Astronomy, University of Minnesota, 115 Union Street SE, Minneapolis, MN 55455, USA

[2]Molecular Engineering & Sciences Institute, University of Washington, Seattle, WA, 98195, USA. jcaroth@uw.edu.

[3]Department of Chemical Engineering, University of Washington, Seattle, WA, 98195, USA. jcaroth@uw.edu.

[4]Center for Synthetic Biology, University of Washington, Seattle, WA, 98195, USA. jcaroth@uw.edu.

***e-mail:** khaki005@umn.edu; noireaux@umn.edu

**ABSTRACT**

CRISPR systems experience off-target effects that interfere with the ability to accurately perform genetic edits. While empirical models predict off-target effects in specific platforms, there is a gap for a wide-ranging mechanistic model of CRISPR systems based on evidence from the essential processes of target recognition. In this work, we present a first principles model supported by many experiments performed *in vivo* and *in vitro*. First, we observe and model the conformational changes and discrete stepped DNA unwinding events of SpCas9-CRISPR's target recognition process in a cell-free transcription-translation system using truncated guide RNAs, confirming structural and FRET microscopy experiments. Second, we implement an energy landscape to describe single mismatch effects observed *in vivo* for spCas9 and Cas12a CRISPR systems. From the mismatch analysis results we uncover that CRISPR class 2 systems maintain their kinetic proofreading mechanism by utilizing intermittent energetic barriers and we show how to leverage the landscape to improve target specificity.

**Keywords:**

CRISPR; Off-targeting effects; Cell-free transcription-translation; Statistical mechanics; Kinetic Proofreading

**Abbreviations:**

TXTL (cell-free transcription-translation); CRISPR (clusters of regularly interspaced short palindromic repeats); eGFP (enhanced green fluorescent protein);

## I. INTRODUCTION

CRISPR technologies are revolutionizing molecular biology by providing highly convenient tools for gene editing [1–4]. The off-target effects caused by insufficient specificity of CRISPR systems, however, are a major bottleneck in genetic engineering [5–7]. The current approaches for predicting such unwanted edits rely on machine learning algorithms trained on large data sets of guide RNAs [8–11]. On the one hand, data-dependent models are highly accurate in predicting the efficiency of a given guide and are ubiquitously used to minimize off-target effects *in vivo*. On the other hand, such models are biased by the experimental conditions in which the data were collected and are not based on mechanistic knowledge of CRISPR targeting, therefore cannot be used to elaborate a deeper understanding. Many experimental results form a qualitative picture of the CRISPR mechanism [1,12,13], yet a coherent quantitative picture is lacking. While successful at grasping some of the CRISPR molecular mechanisms, such as hybridization [14,15], PAM search [16], and cleavage dynamics [16,17], the current biophysical models based on statistical mechanics do not completely capture the anomalous off-targeting effects observed in the experiments [11,18–20]. We propose a model of CRISPR systems that both explains off-targeting effects and proposes rational improvements to the accuracy of CRISPR technologies consolidating our comprehension of CRISPR systems.

Our model includes the following mechanisms in the CRISPR target recognition process: PAM binding, DNA unwinding, RNA-DNA hybridization, and conformational changes of the Cas enzymes. These steps have been evidenced by structural studies [12,21–24] and by FRET microscopy [25–29]. We demonstrate that the RNA-DNA hybridization proceeds by discrete un-windings of target DNA that correspond to the conformational changes of the spCas9 enzyme. Keeping the guide RNAs at energetically favorable lengths enables performing CRISPR silencing and activation with spCas9 enzymes. We convert the energy landscapes inferred from cell-free experiments into a previously developed birth-death process model [14], which allows explaining accurately the anomalous indel rates seen in spCas9 and Cas12a experiments [19,20,30,31]. We extend the model to describe the single mismatch patterns in CRISPR Cas3 and Cas13 [15,32]. With further extrapolation, our model illustrates that periodic mismatch tolerances are a product of the intermittent energetic barriers, which create a kinetic proofreading like mechanism [33] in CRISPR systems.

## II. Results and Discussion

### A. Energy landscape

The cell-free transcription-translation (TXTL) system used in this work [34,35] enables characterizing CRISPR technologies [36], with an excellent agreement between the observations made *in vitro* and *in vivo*. We harnessed the specific advantages of CRISPRi (CRISPR interference) [37,38] and CRISPRa (CRISPR activation) [39] systems in TXTL experiments to determine the energy landscape of CRISPR-spCas9 **(Fig. 1)**. By pairing CRISPRi and CRISPRa, and using both active and deactivated enzymes, we can compare their signatures and obtain a quantitative image of CRISPR spCas9's conformational changes in TXTL.

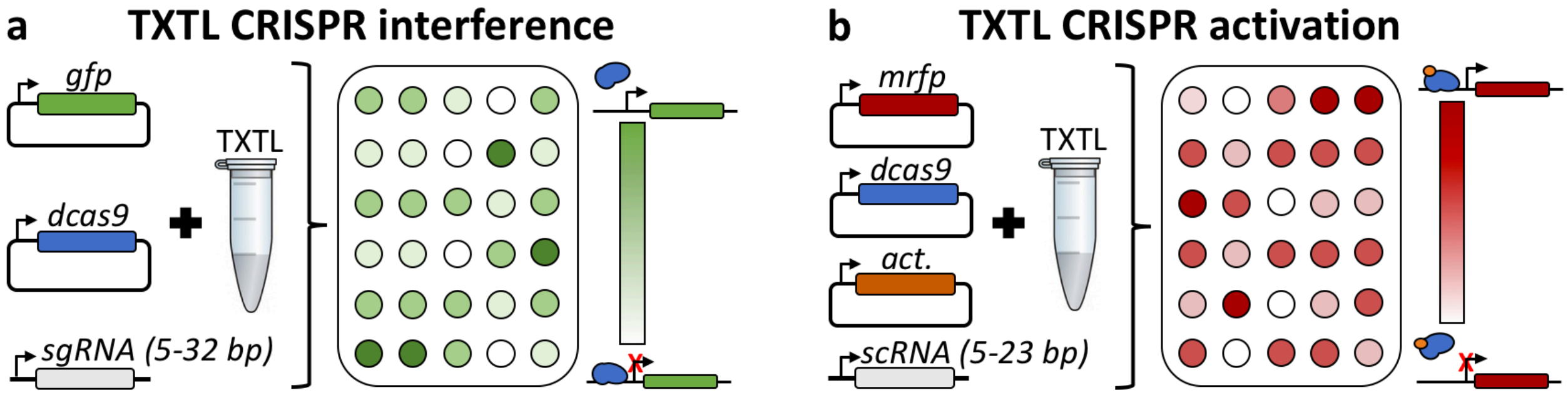

**Fig. 1.** Schematic of the TXTL experiment. **(a)** The TXTL CRISPR interference experiment is composed of two plasmids to express the genes *egfp* and *spdcas9*, and a linear DNA to express the *sgRNA*. The sgRNA targets the promoter of the *egfp* gene. The reaction is assembled, incubated at 29 °C on a well plate and fluorescence signals are analyzed quantitatively using either kinetics or endpoints measurements. CRISPRi experiments work best for the exact measurement of the fraction of targets bound by CRISPR enzymes. However, the silencing effects caused by a spdCas9 enzyme are difficult to distinguish from the cutting effects of spCas9,

therefore tracking the catalytic activity with CRISPRi is difficult. **(b)** The TXTL CRISPR activation is composed of three plasmids to express the genes *mrfp*, *spdcas9* and *activator* (labeled as *act*), and a linear DNA to express the *scRNA*. The scRNA targets the promoter of the *mrfp* gene. The reaction is assembled, incubated at 29 °C on a well plate and fluorescence signals are analyzed quantitatively using either kinetics or endpoints measurements. With CRISPRa experiments we cannot accurately measure what fraction of targets are occupied by an enzyme, because the measured signal is the expression of the target. However, CRISPRa enables visualizing the transition from nuclease inactive to nuclease active spCas9, because a catalytically active enzyme cuts and thus silences the target rather than activating it. Therefore, with both CRISRPi and CRISPRa we can observe the locations of discreet unwinding events and conformational changes.

We determined the energy landscape of CRISPR-spCas9 systems' target recognition by performing a series of TXTL reactions with varying lengths of sgRNAs for CRISPRi and scRNAs for CRISPRa. In both cases, we observe an anomalous strong binding for guide RNA lengths much smaller than the conventional optimal length around 20 bp **(Fig. 2a and Fig. 2b)**. The anomalous peaks can be reduced in comparison to the 20 bp-long guide RNA when the concentration of added guide RNAs is changed **(Fig. 2c and Fig. 2d)**. To understand these results we model the binding of CRISPR-RNA complexes to DNA as a Michaelis-Menten process:

$$[CRISPR] + [Target\ Free] \leftrightarrows [Target\ Bound] \qquad (1)$$

We assumed that the association rate constant $k_{on}$ is constant for a guide RNA series because it is controlled by the diffusion of the CRISPR enzyme and the binding to the PAM sequence. The PAM sequence was identical for all the sgRNAs or scRNAs in a series of target length, and diffusion remained identical for all the experiments. However, it is possible that the length of the scRNAs/sgRNAs strongly affects both the complex formation efficiency and the association rate and should be considered for more involved analysis. For bacterial CRISPR experiments the dissociation rate of CRISPR complexes with a fully matched guide from the target is controlled mainly by DNA replication and the unassisted unbinding events are negligible. DNA replication does not occur in TXTL, therefore the unassisted unbinding events are the main contribution to the dissociation rate in TXTL. Using the Arrhenius equation we express the dissociation rate as $k_{off} = k_0 \exp(-\beta^* E_{binding})$, where $k_0$ is the off rate of a Cas enzyme with a short and completely mismatched guide RNA. Consequently, we can express the equilibrium probability of forming a target-CRISPR complex as:

$$p_b = \frac{1}{1 + \alpha\, e^{(-\beta E_{binding})}}, \quad \alpha = \frac{K_{d0}}{[CRISPR]} \qquad (2)$$

where $K_{d0}$ is the equilibrium dissociation constant of a short and fully orthogonal guide RNA and *[CRISPR]* is the concentration of the CRISPR-RNA complex. Using equation (2) we determined an energy landscape of effective binding energy with respect to the length of the guide RNA. We hypothesize that the intermittent decreases in binding energy are caused by discrete DNA un-windings controlled by conformational changes in the Cas enzyme. Previous experiments demonstrated that conformational changes of CRISPR enzymes also occur in discrete steps that correspond to sequential hybridization and acts as conformational checkpoints in target recognition [25,26]. To experimentally confirm that un-windings correspond with conformational changes, we replace spdCas9 in the CRISPRa experiment [40] with spCas9, thus allowing the system to perform cuts when the Cas enzyme becomes catalytically active. With this substitution gene expression is activated with scRNA lengths below 15 bp. At greater lengths the CRISPR-spCas9 system cuts the target, thus reducing the expression to background levels **(Fig. 2e)**. The length of the scRNA at which we observe the conformational from inactive to active corresponds to the length at which the effective binding energy decreases in spCas9 CRISPRa experiments **(Fig. 2b).** This result indicates that conformational shifts correspond to the discrete hybridization of the guides. We notice that CRISPR systems have the first peak and thus first checkpoint at the 8-12 bp **(Fig. 2c, 2d, S1c, S1d, S1e, S1f and S1g)**, which corresponds to the widely reported length known as the seed region [1]. The conformational shift from catalytically inactive to active for spCas9 corresponds to lengths of 16-18 bp [25,40]. While a 20 bp guide RNA is enough to cut, further checkpoints are seen as we keep increasing the guide length **(Fig. 2a, Fig. S2f)**, which indicates that perhaps there are further conformational changes to the spCas9 enzyme. The final conformational state a CRISPR system can enter is thus dependent on the length of the target sequence of guide that form the complex with the Cas enzyme. It is unclear what effect the steric hindrance of the activator has on the guide length dependence in CRISPRa experiments. It is possible that the conformational locations and energies differ for an activator-free CRISPR binding event. A concentration of 4nM of dCas9 plasmid was added to the dCas9 reactions and 3nM of Cas9 plasmid was added to the Cas9 reactions

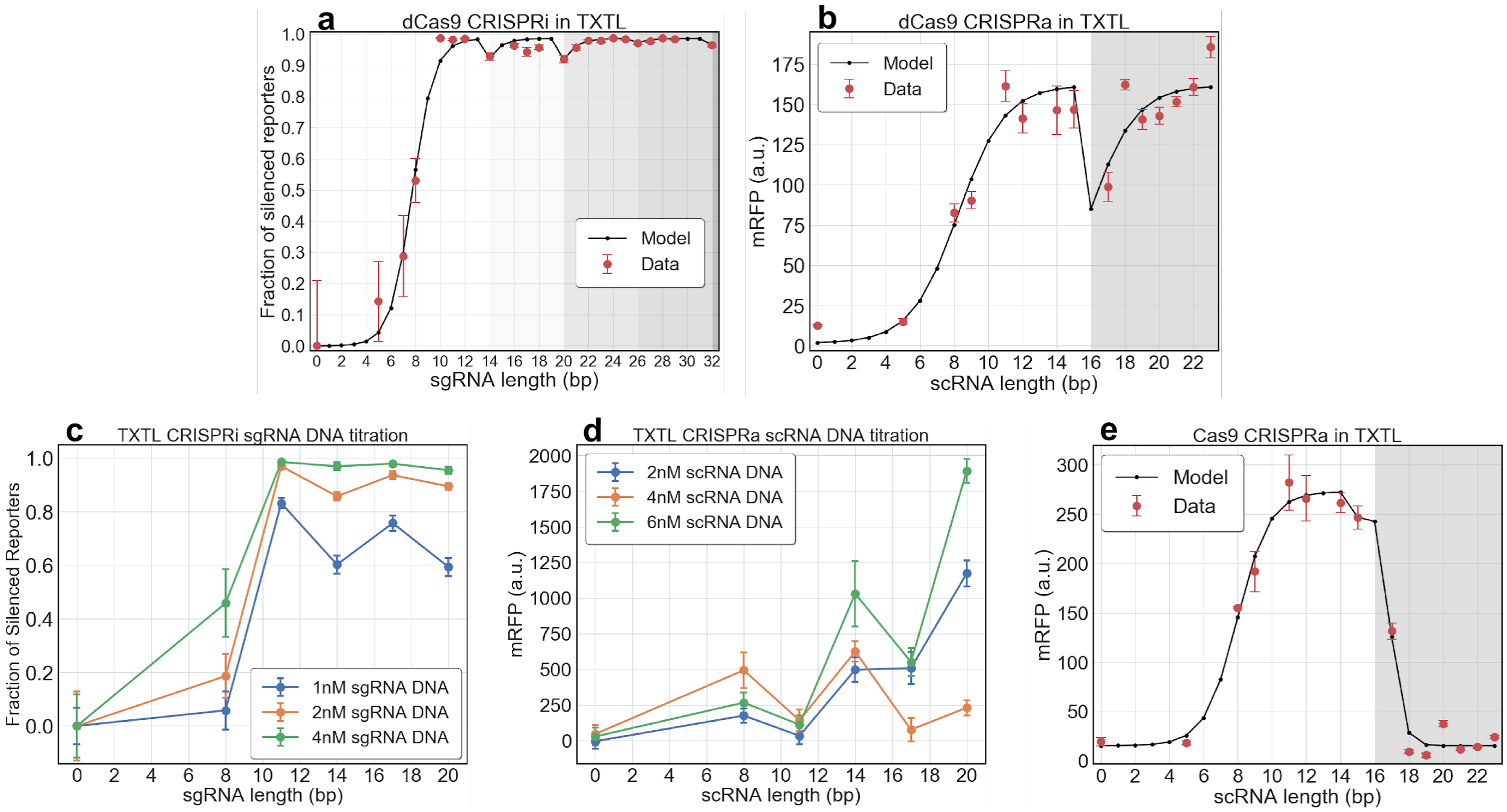

**Fig. 2.** CRISPRi and CRISPRa TXTL experiments for varying sgRNAs and scRNAs. The different background shades show the different conformational states. **(a)** CRISPRi with different sgRNA target sequence lengths in TXTL. The ratio of bound targets tends to increase with the larger sgRNA target sequence lengths, except for base pairs 14, 17, 20, 26, and 32. These drops occur due to energetically costly intermittent unwinding events of the target DNA and the conformational changes corresponding to those un-windings. **(b)** CRISPRa with different scRNA target sequence lengths in TXTL. In this experiment we observe a drop with a rebound in the mRFP for a 17bp long target sequence for a spdCas9-CRISPRa experiment. **(c) and (d)** CRISPRi/CRISPRa in TXTL with varying concentrations of sgRNA/scRNA DNA and the target lengths. The concentration of the scRNA and sgRNA determines the number of CRISPR complexes in the reaction. **(e)**, TXTL CRISPRa experiment performed with the spCas9 enzyme. Protein synthesis drops to background level for targets longer that 17 bp.

Based on the results from TXTL experiments we can make a model for the energy landscape of spCas9 CRISPR target recognition **(Fig. 3)**. Depending on the length of target sequence on the guide RNA, the CRISPR complex will be able to enter 3 different conformational states: seed, pre-active, active [38]. Similar energy landscapes have been observed in experiments that directly look at DNA unwinding [41]. If the guide length is shorter than the seed transition length, then the CRISPR enzyme acquires binding energy through hybridization with the target DNA. However, such a guide is too short for the CRISPR enzyme to make an energetically costly transition to the pre-active state. As a result the binding probability has a peak at lengths that are below the required lengths for transition. In contrast, guides with lengths 1-3 bp larger than the pre-active transition length have weak binding. This is caused by the inability of such guides to energetically compensate for the discrete unwinding of the pre-active region nucleotides. Because the energetic cost of entering the state is larger than the energy available due to RNA-DNA hybridization, unbinding is a more likely outcome. Similar effects occur for guides that are 1-3 bp longer than the length needed for the active conformational state transition.

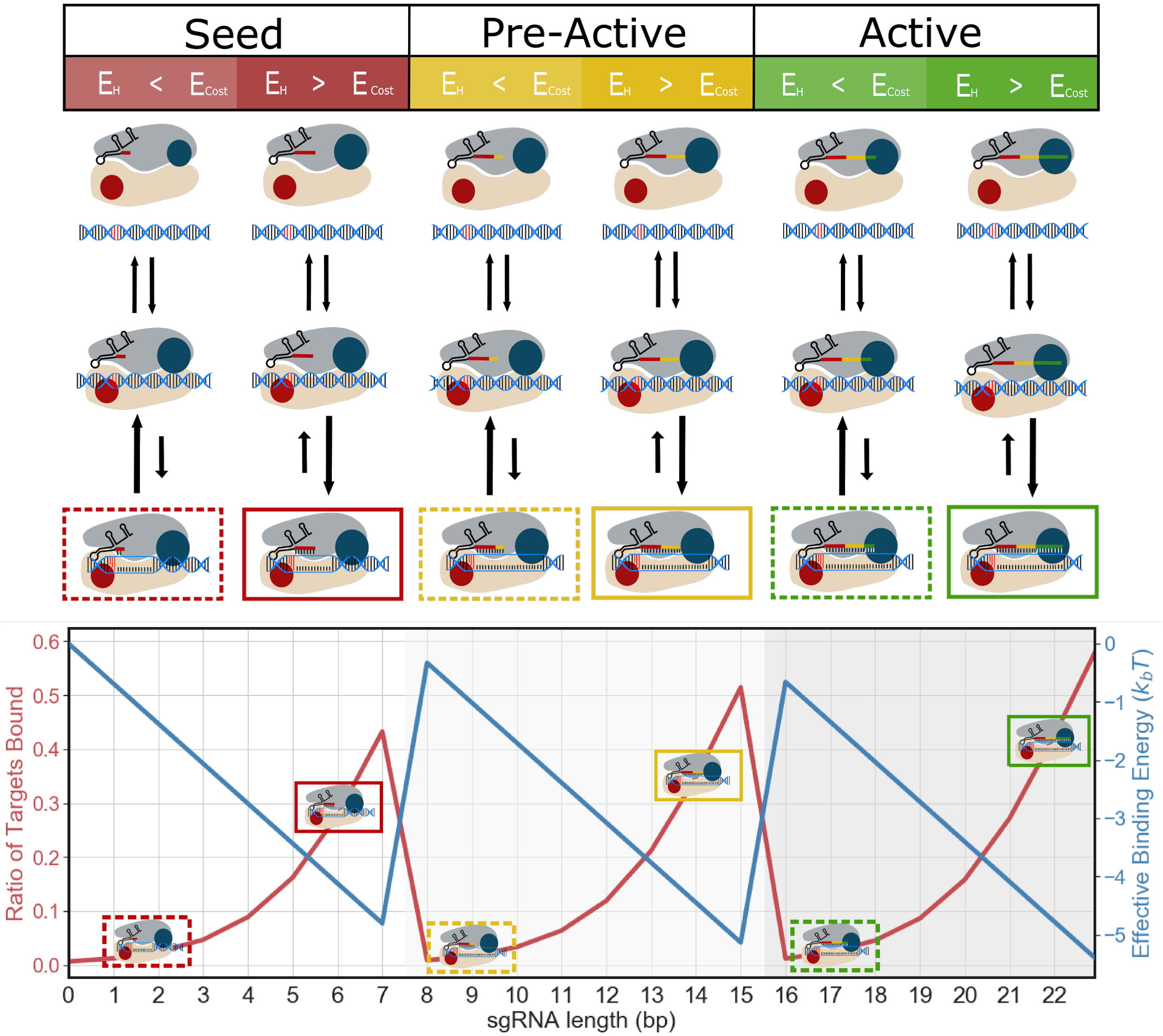

**Fig. 3.** spdCas9-CRISPR target recognition depending on the length of sequences. The recognition process of CRISPR systems can be split into 3 conformations: seed, pre-active, active. Each conformation has a length at which the CRISPR complex is likely to transition to that conformation and an energetic cost for making that transition. In the schematic the transition to the pre-active conformation occurs at the 8th nucleotide and the transition to the active

conformation occurs at the 16th nucleotide. $E_H$ is measured by the number of nucleotides available for hybridization, while $E_{Cost}$ is measured by the cumulative costs of transitions to reach the state which the guide length corresponds to. If the number of base pairs beyond the transition is large enough to compensate for the energetic costs of the transitions then the CRISPR enzyme is more likely to remain bound. Examples of weak binders in each state are shown in dashed boxes. In weak binders guides are 1-3bp longer than the length requirement to enter the state, therefore $E_H$ is less than $E_{Cost}$. Examples of strong binders are shown in the solid boxes. CRISPR systems equipped with guides that have targets 1-3bp shorter than the length needed for a transition to the next conformation hybridize enough nucleotides to compensate the conformational costs and therefore $E_H$ is larger than $E_{Cost}$.

**B. Mismatch analysis**

We implement the energy landscapes determined by TXTL experiments towards explaining off-targeting effects observed for *in vivo* experiments. The landscape gains $E_{pam}$ for every PAM site bound and gains $E_c$ for a nucleotide match. The landscape loses $E_{mm}$ for a nucleotide mismatch. The energy for cutting the DNA is $E_{cut}$. The conformational shift costs are $E_{cs,i}$ and are located after the $n_{cs,i}$ nucleotide. We model (i) spCas9 with two conformational shifts **(Fig. 4a)**, and (ii) Cas12a with one conformational shift **(Fig. 4b)**. The energies were constrained to values reported previously [42,43].

CRISPR proofreading has been previously modeled as a birth-death process (BDP) across an energy landscape [14]. Each state of the BDP corresponds to a conformation of the CRISPR-DNA complex in which a number of RNA-DNA bonds have formed. To simplify the model it is assumed that transitions are available only between nearest neighbor states and that the transition rates are independent of the path previously covered by the Cas enzyme. We can then describe each state of the process with a partition function that includes the energy of the state and of its nearest neighbors:

$$z_i = 1 + e^{(-\beta(E_{i+1}-E_i))} + e^{(-\beta(E_{i-1}-E_i))} \tag{3}$$

The transition probabilities for an arbitrary unit of time are then:

$$p_i = \frac{e^{(-\beta(E_{i+1}-E_i))}}{z_i} \tag{4}$$

$$q_i = \frac{e^{(-\beta(E_{i-1}-E_i))}}{z_i} \tag{5}$$

where $p_i$ and $q_i$ are forward and reverse motion respectively. The process also has two absorbing states: CRISPR unbinding from the DNA and CRISPR cutting the DNA. The unbinding transition occurs when the DNA-protein interaction at the PAM site is disrupted. The target can be cut from any of the states located in the catalytically active part of the landscape as seen in truncation experiments. The cutting probability is determined by a simple two-state partition function containing the energy needed to cut and the energy to remain uncut. Therefore the probability of cutting is:

$$r = 1 - (1 + e^{(-\beta E_{cut})})^{-1} \tag{6}$$

With all the transition rates of the process known we can assemble a transition matrix *T* and evolve the system probability density ***S(t)***:

$$S(N) = T^N S(0) \tag{7}$$

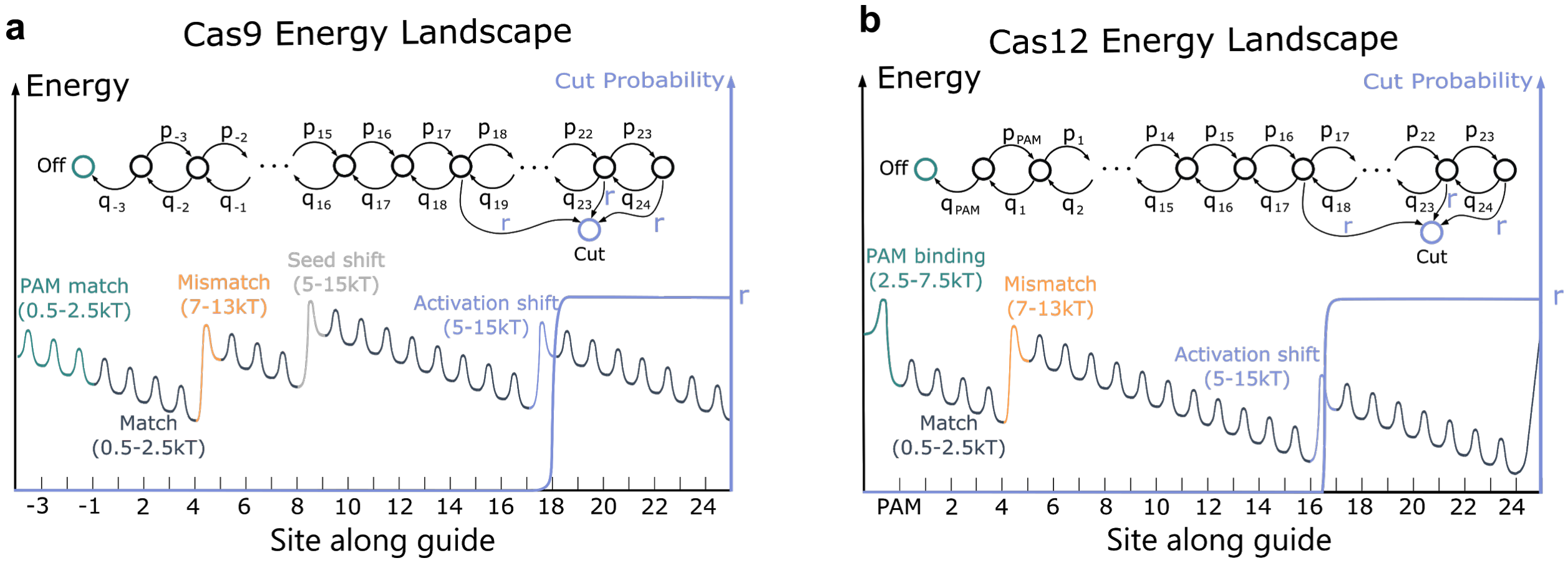

**Fig. 4.** CRISPR kinetic recognition model. **(a)** Representation of an energy landscape in CRISPR spCas9 target recognition (black) and its respective cutting probability (blue) as functions of the site along the guide RNA. As seen in TXTL experiments spCas9 is modeled with 2 conformational shifts: seed shift and activation shift. The landscape is translated to a birth-death process with rates $p_i$ and $q_i$ for motion along the guide and r for cutting the target. **(b)** Representation of an energy landscape in CRISPR-Cas12 target recognition (black) and its respective cutting probability (blue) as functions of the site along the guide RNA. The best mismatch results were achieved with a single activation shift. The landscapes are translated to a birth-death process with rates $p_i$ and $q_i$ for motion along the guide and r for cutting the target for both spCas9 and Cas12 target recognition.

To decrease the computational load of the model **(Fig. 5a)**, we renormalize the transition matrix of the process into submatrices that encapsulate transitions between conformational states instead of hybridization number states **(Fig. 5b and 5c)**. The submatrices are picked such that the absorbing states of a submatrix are the initial states of the adjacent submatrices. To determine the rates of the renormalized Markov chain **(Fig. 5d)**, we acquire the probabilities of being located in the absorbing states of each submatrix after evolving the initial state with the submatrix *M* until the distribution is focused in the absorbing states:

$$M^N|initial\ state\rangle = rev|absorbing\ state\ rev\rangle + forw|absorbing\ state\ forw\rangle \tag{8}$$

Because the dynamics of a single path is independent of the dynamics of all possible paths, such renormalization will only cause a disruption in the dynamics of the process, but will not affect the final probabilities. We evolve the renormalized Markov chain until all probabilities are concentrated in the Unbound and Cut states. Therefore we produce the same final probabilities as we would with the full Markov chain, but with less computational cost.

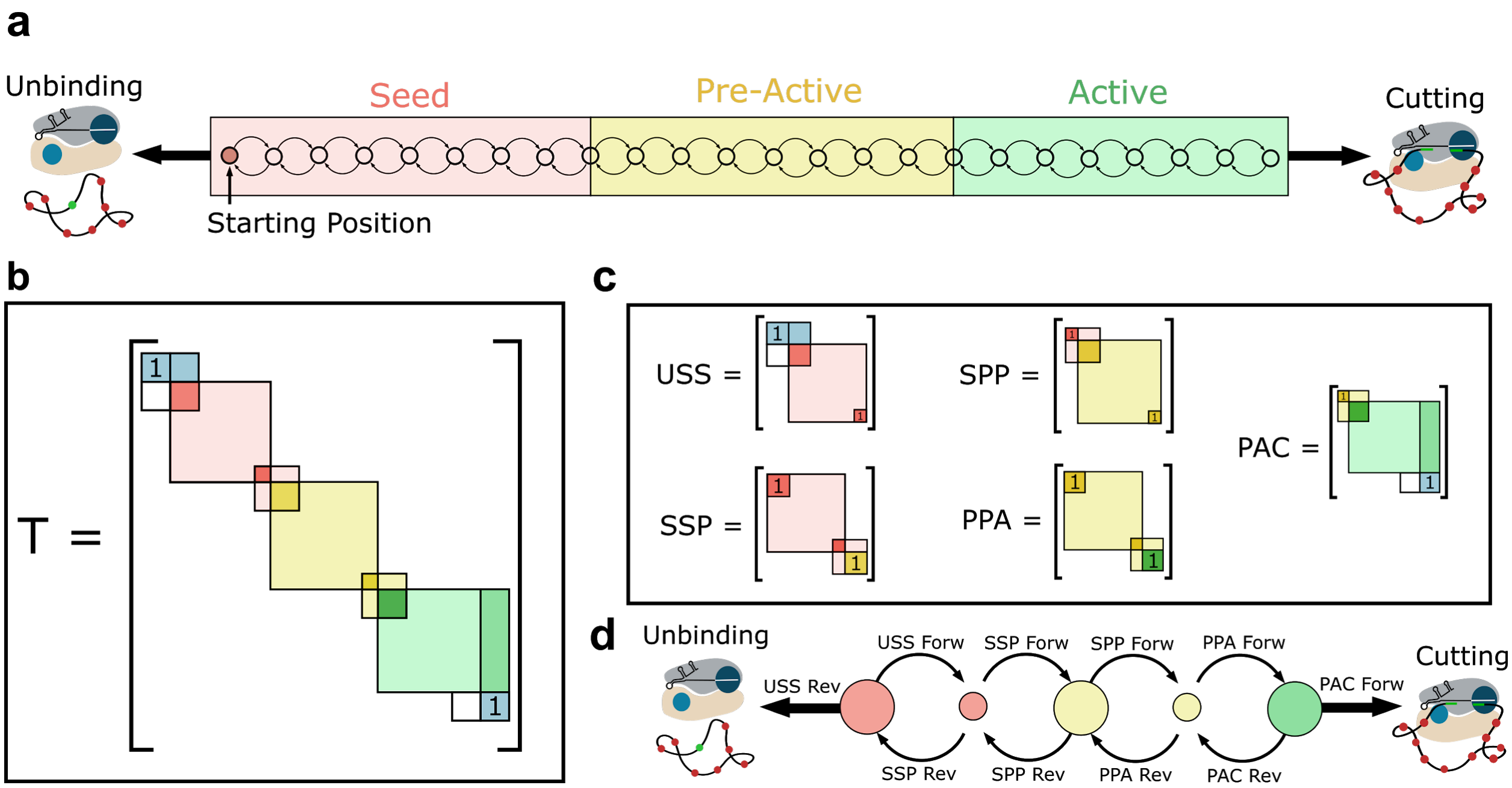

**Fig. 5.** Absorbing Birth-Death process renormalization. **(a)** The schematic of the original hybridization Markov chain. Each node is a hybridization state, while the colors signify the conformational states belonging to the set of hybridization states. **(b)** The transition matrix of the hybridization process. The colors correspond to the colors in the Markov chain. **(c)** Transition submatrices that define the rates of the reduced Markov chain. The letters of the submatrices stand for the conformational states that are present in the matrix. The first and the last letters are the absorbing states and the middle letter is the starting location (USS: Unbinding - Seed - Seed, SSP: Seed - Seed - Pre-active, SPP: Seed - Pre-active - Pre-active, PPA: Pre-active - Pre-active - Active, PAC: Pre-active - Active - Cut) . The PAC matrix demonstrates that cutting can occur from any hybridization state in the active region, therefore the rightmost column is filled. **(d)** The schematic of the renormalized Markov chain formed from converting the coefficients of the absorbing states of the submatrices into the rates. The seed and pre-active regions have two nodes that correspond to passing the region and transitioning to the next region respectively. The active region is described by a single node, because cutting is the final state.

When we fit the spCas9 *in vivo* mismatch experiments [19], we observe the first conformational shift at the 8-9 bp and the second conformational shift at the 15-16 bp, which agrees with the TXTL experiments **(Fig. 6a and 6b)**. For the Cas12a *in vivo* experiments [20,31], we observe a single conformational shift at 17-18 bp **(Fig. 6c and 6d)** [27,28]. The average hybridization energies per base pair for spCas9 fits are 1.38 $k_bT$ and 1.98 $k_bT$ for VCP2 and PSMD7 targets respectively. The average hybridization energies per base pair for AsCas12a/AsCpf1 targeting DNMT1-1 and LbCas12a/LbCpf1 targeting DNMT1-3 are 1.17 $k_bT$ and 1.01 $k_bT$ respectively. This result agrees with the results from the previous kinetic landscape model [14] and with experimental results on the binding energies of RNA-DNA bonds [43]. The mismatch penalties in the spCas9 fits are 7.56 $k_bT$ and 9.57 $k_bT$ for VCP2 and PSMD7 targets respectively. The average hybridization energies for AsCas12a/AsCpf1 targeting DNMT1-1 and LbCas12a/LbCpf1 targeting DNMT1-3 are 8.19 $k_bT$ and 8.77 $k_bT$ respectively. The expected penalty of an average mismatch is approximately the cost of two hydrogen bonds, which is observed in our results considering an average hydrogen bond of interest is between 3-5 $k_bT$ [44]. For Cas9 fitting, the PAM consists of 3 successive bonds each with an energy of 1.98 $k_bT$ and 1.62 $k_bT$ for VCP2 and PSMD7 targets respectively, thus making the total PAM energy larger than the average hybridization energy per nucleotide, as described in the requirements of the previous model that PAM binding needs to be significantly stronger than the energy gained by a hybridization step [14]. Cas12 PAMs were modeled as a single binding event with bond energies of 2.61 $k_bT$ and 2.56 $k_bT$ for AsCas12a/AsCpf1 targeting DNMT1-1 and LbCas12a/LbCpf1 targeting DNMT1-3 respectively, therefore the Cas12 values also satisfy the requirement. The energy values of the first conformational change are 10.79 $k_bT$ for the VCP2 target and 10.26 $k_bT$ for PSMD7 target. The energies of the catalytically activating conformational change are 10.46 $k_bT$ for the VCP2 target and 13.01 $k_bT$ for PSMD7 target. The energies of the activating conformational changes in Cas12 are 11.92 $k_bT$ for AsCas12a/AsCpf1 targeting DNMT1-1 and 10.29 $k_bT$ for LbCas12a/LbCpf1 targeting DNMT1-3 respectively.

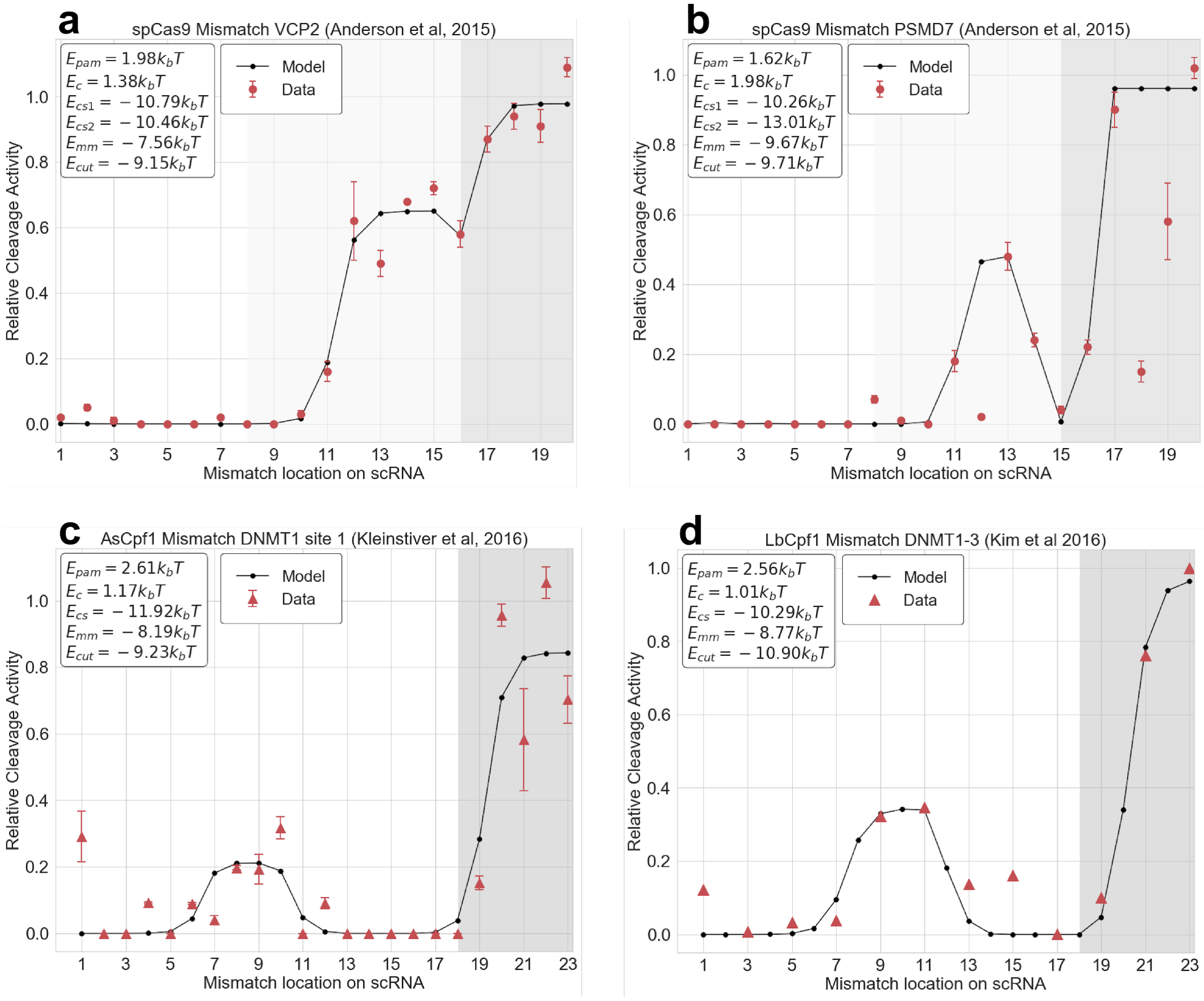

**Fig 6.** Kinetic Landscape Analysis. **(a) and (b)** Fits of the kinetic landscape model to single mismatch spCas9 data targeting VCP2 and PSMD7 respectively. The different colors represent different conformational regions of spCas9. **(c) and (d)** Fits of the kinetic landscape model to single mismatch LbCas12a and AsCas12a respectively. As in spCas9, the colors represent the conformations. The energies of the landscape used for fitting are described in the legends of the plots. The locations of the conformational changes are shown in the background color transitions.

In spCas9 target recognition the first conformational shift serves as a checkpoint and ends the seed region of spCas9. spCas9 accumulates binding energy in the seed region, therefore mismatches in that region favors spCas9 towards unbinding. Placing mismatches in the pre-active region tends to be more tolerable, because this region is reinforced by a fully matched seed. Even if a mismatch in the pre-active region reverses the state of CRISPR back to the seed state, it is likely to progress again into the pre-active region. Mismatches in the active region of spCas9 are the most tolerable, because the hybridization of a single base pair would facilitate enough stability to perform edits **(Fig. 7a)**. In Cas12 experiments, to explain the anomalous mid-gRNA mismatch tolerance, we look at a coarse-grained picture of the energy landscape **(Fig. 7b)**. A mismatch divides the seed/pre-active state into two energetic wells **(Fig. 7c)**. If a mismatch is placed early, the first well is biased to reverse, while the second well is biased to move forward. If a mismatch is placed at the end of the region, the first well facilitates movement into the second well, but the second well is now biased to reverse. The reverse motion dominates the dynamics in those areas of the guide, meaning that mismatches early and late in the seed/pre-active region are less tolerable in Cas12a. However, for a mismatch in the middle of the seed region, both wells are biased towards forward movement, therefore making such mismatches tolerable. Mismatches in the active region of Cas12a are the most tolerable, because the catalytic activity is independent of the active region hybridization and double mismatches at the end of the guide RNA do not strongly affect the nuclease efficiency of Cas12 enzymes **(Fig. S2b)**.

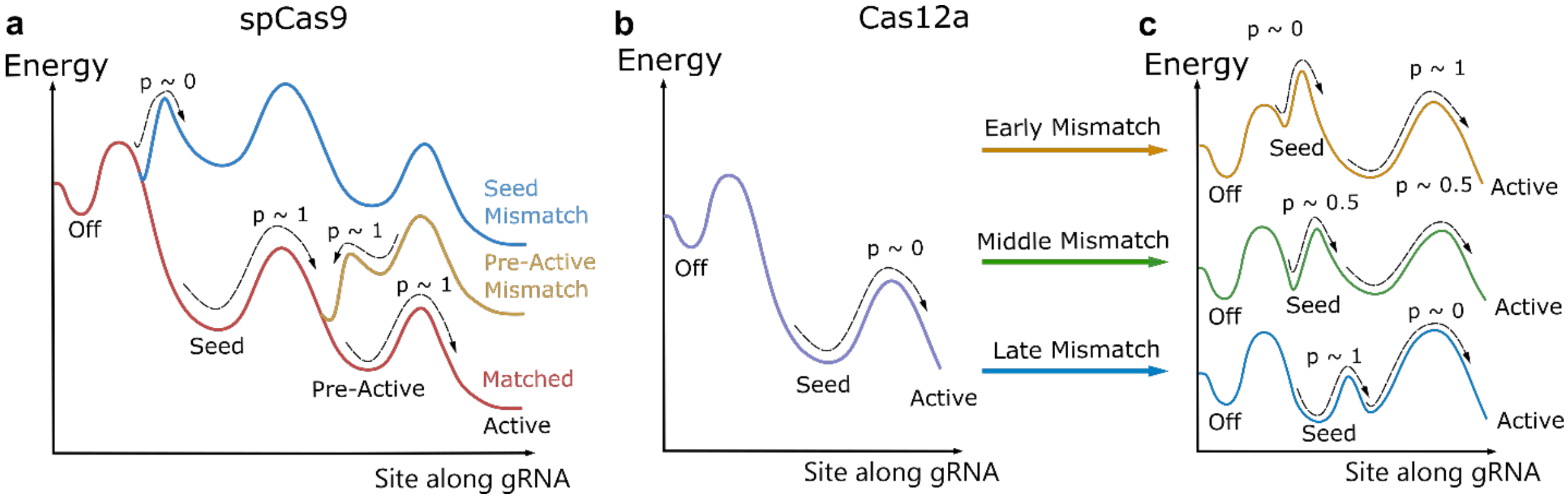

**Fig. 7.** Graphical explanation of mismatch tolerance in CRISPR systems. **(a)** Coarse-grained representation of a CRISPR-spCas9 energy landscape. For a fully complementary guide (red), the probability of transitioning to the active region is highly likely. If a mismatch is placed in the seed region of the landscape (blue), the probability of reaching the active state is low because the energy acquired before the mismatch is less than the mismatch penalty energy. If a mismatch is placed in the pre-active region, the transition from the pre-active state to the active state is unlikely, but because the seed is fully matched the system is likely to return to the pre-active state. Therefore, the probability current is contested between the mismatch and the fully matched seed and depending on the location of the mismatch the off-targets effects are different. **(b)** Coarse-grained energy landscape of Cas12 focusing on the seed region. The arrow demonstrates the likelihood of progressing through the seed into the pre-active state during. **(c)** Coarse-grained energy landscapes with the introduction of mismatches at different locations. The arrows demonstrate the approximate likelihoods of progressing from one local energetic minima to the other. For early mismatches the probability of passing the first effective well is near zero, while for the late mismatches the probability of passing the second effective well is near zero, therefore for both of the mismatches, reaching the next conformation is unlikely, while for the well place mismatch, both probabilities are non-zero.

The same analysis can be applied to Cas3 and Cas13 CRISPR systems, despite the significant differences in the mechanism **(Fig. S2c, S2d, S2e)**. Cas13 recognizes RNA targets instead of dsDNA targets, therefore there are no unwinding steps in the recognition mechanism. However, structural analysis reveals conformational changes during target recognition that change the binding availability of the guide [13,32] and these conformational changes are utilized by the Cas13 enzyme as the proofreading checkpoints. Also, because Cas13 cuts RNA non-specifically once it is catalytically active, the effect an active enzyme has on the population of targets is different from the effect described in our model.

Cas3 CRISPR consists of multiple enzymes that form a complex, including a helicase enzyme. Cas3 utilizes ATP to unwind the target, unlike spCas9 and Cas12 which rely on thermal fluctuations. When fitting the model to available Cas3 mismatch data [15], we notice that the periodical mismatch insensitivity can be explained by the helicase unwinding the target in discrete 6-bp steps **(Fig. S2b)**. The energetic costs of these un-windings are 2-4 $k_bT$, which is smaller and more consistent than the costs seen in spCas9 and Cas12 that range between 10-14 $k_bT$.

### C. Energetic constraints, speed-accuracy trade-offs, computational possibilities

The introduction of conformational changes maintains states at energetic wells with depths comparable to mismatch energies **(Fig. 8a)**. This strategy increases the specificity of Cas enzymes because a larger fraction of single mismatches create statistically insurmountable energetic landscapes. The specificity of CRISPR systems can be further improved (i) by decreasing the average hybridization energy of RNA-DNA bonds or increasing the average hybridization of DNA-DNA bonds **(Fig. 8b)**, (ii) by increasing the energetic costs of conformational changes **(Fig. 8c)**, and (iii) by increasing the energetic cost of cutting **(Fig. 8d)**.

Previous works have optimized the performance of CRISPR enzymes by changing the protein sequence in locations critical for target recognition [25]. However, we observe that when the accuracy of the recognition process is improved by scaling the conformational change costs, the system experiences a loss in the speed of recognition **(Fig. 8e)**. Therefore for any biotechnological application of CRISPR we cannot have both relatively high accuracy and relatively high processing speed using the CRISPR recognition mechanism. In nature, CRISPR enzymes need to process their targets in time frames comparable to viral replication, therefore evolutionarily natural Cas enzymes are optimized on the basis of both speed and accuracy. For CRISPR gene, editing accuracy is of higher value, thus sacrificing speed in exchange for higher specificity is recommended. The engineering of tailored CRISPR systems can be done through directed evolution [31,42] or by mining for new CRISPR systems that will have more thorough target recognition mechanisms [45].

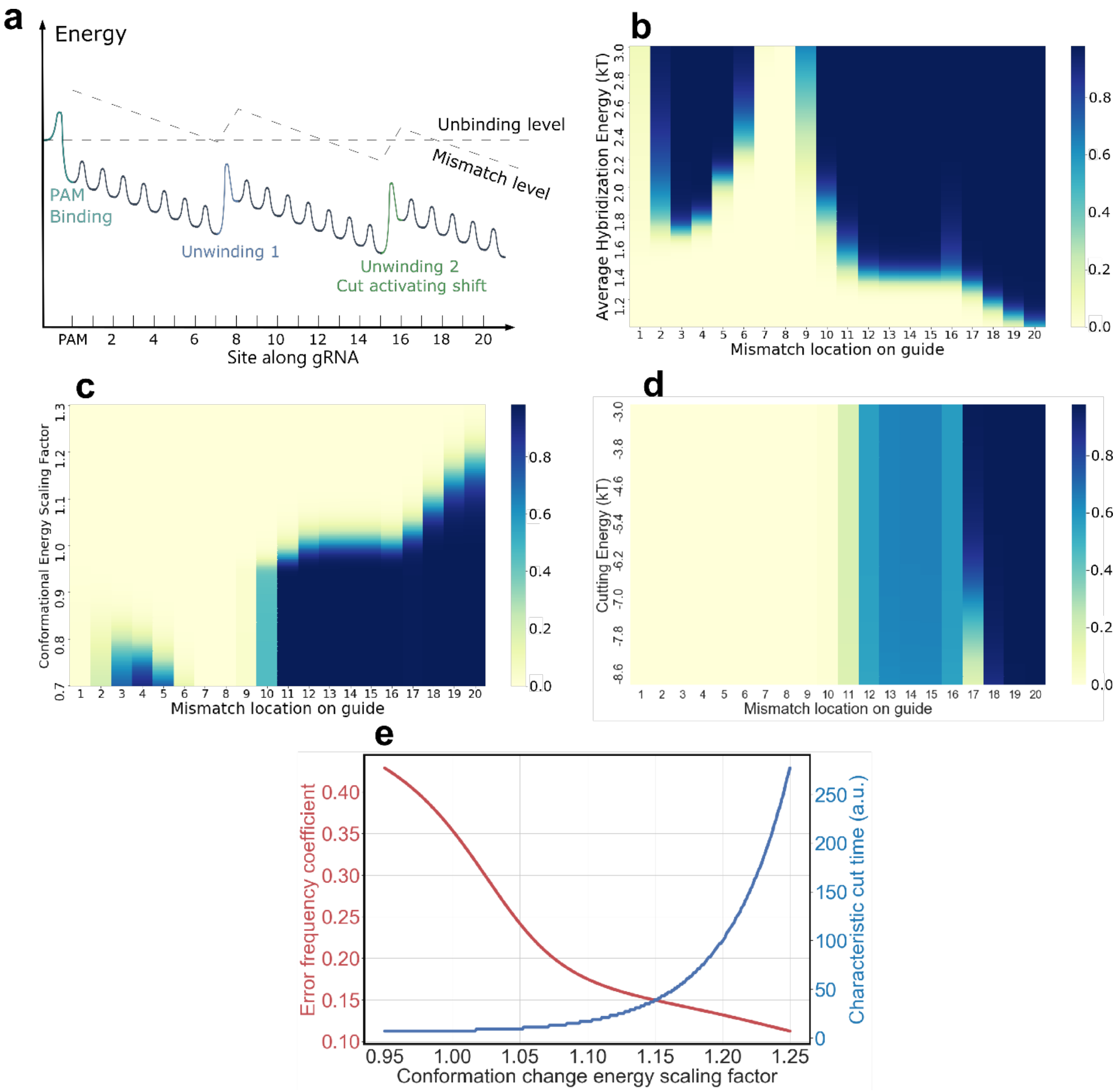

**Fig. 8.** Improving accuracy of spCas9 CRISPR. **(a),** the energy landscape of spCas9 CRISPR contains two checkpoints before reaching the active state. The mismatch level demonstrates the height of an energy landscape if a mismatch were to occur at any given site. If the height is larger than the energetic level of unbinding, then the system is biased to unbind. **(b),** Changing the average hybridization energy of the energy landscape affects the specificity of the guide. Weaker guides tend to have less off-targeting under the condition that all other parameters are independent of the average hybridization energy. **(c),** Adjusting the energetic costs of unwinding events/conformational changes by a linear factor also affects the specificity of an arbitrary guide. Increasing the size of the checkpoints increases the specificity of the recognition process. **(d),** Changing the energy required to cut the target affects the specificity of the base pairs in the active region. Increasing the cutting energetic requirements makes the cutting more specific. **(e),** the speed-accuracy trade-off in CRISPR-like target processing. The error frequency coefficient is the tolerance of an average mismatch in the guide and the characteristic cut time is the time required for $1\text{-}e^{-1}$ of possible paths to reach the cut state

## III. Conclusion

The application of simple models based on the energetic landscapes is a viable strategy for gaining quantitative insight of CRISPR systems. While our approach allows for a quantitative analysis, our model is not fully comprehensive. To improve our results, it is important to develop models that consider the full system, where the environment of the cell plays a larger role. Our model neglects the dynamics of CRISPR during target recognition. Quantitative understanding of how truncations and mismatches affect the recognition process temporally is critical for understanding the underlying molecular principles behind CRISPR target recognition systems. Finally, an important task for first-principles genetic engineering is uncovering the connection between the sequences and the conformational energies of CRISPR. In this work we have demonstrated a connection between the off-targeting profile of an arbitrary sequence with its energy landscape. Further work in this direction will measure how sequences determine the energy landscape of the system. As of now we can only perform post-experimental analysis, but with the mentioned models we can improve towards rational prediction in genetic editing.

**Acknowledgments**
J.C. and V.N. acknowledges funding support from the National Science Foundation award CBET 1844152 and the Binational Science Foundation award 2018208. A.K. and V.N. would like to thank Dr. Emily Anderson and Dr. Anja Smith for providing spCas9 mismatch data, Prof. Benjamin Kleinstiver for providing Cpf1 mismatch data, Dr. Omar Abudayyeh, Dr. Jonathan Gootenberg, Dr. Ian Slaymaker, and Prof. Feng Zhang for providing Cas13 mismatch data, Prof. Finkelstein for providing Cas3 mismatch data. A.K. would like to thank Seth Thompson for plasmid amplifications, Diego Alba Burbano for his helpful manuscript comments, Joe Meese and Jacob Ritz for fruitful conversations about the CRISPR models,

**Author Contributions**
A.K., D.G. and V.N. designed the research, A.K. performed the experiments, B.T. and J.F. engineered components for CRISPRa, A.K. and V.N. analyzed the data and wrote the manuscript with input from all of the authors.

**Additional Information**
Supplementary information accompanies this paper.

**ORCID**
Vincent Noireaux https://orcid.org/0000-0002-5213-273X

**Competing Interests**
The Noireaux laboratory receives research funds from Arbor Biosciences, a distributor of the myTXTL cell-free protein synthesis kit.

# Supplementary information

## Proofreading mechanism of Class 2 CRISPR-Cas systems

Aset Khakimzhan[1,*], David Garenne[1], Benjamin I. Tickman[2], Jason Fontana[2], James Carothers[2,3,4], Vincent Noireaux[1,*]

[1]School of Physics and Astronomy, University of Minnesota, 115 Union Street SE, Minneapolis, MN 55455, USA

[2]Molecular Engineering & Sciences Institute, University of Washington, Seattle, WA, 98195, USA. jcaroth@uw.edu.

[3]Department of Chemical Engineering, University of Washington, Seattle, WA, 98195, USA. jcaroth@uw.edu.

[4]Center for Synthetic Biology, University of Washington, Seattle, WA, 98195, USA. jcaroth@uw.edu.

***e-mail:** khaki005@umn.edu; noireaux@umn.edu

**METHODS**

**Materials.** DNA was purchased from Integrated DNA Technologies (Coralville, IA). Unless otherwise mentioned all the other reagents were purchased from Sigma Aldrich (St. Louis, MO).

**DNA constructs.** For silencing experiments in TXTL, the spCas9 or spdCas9 enzymes were synthesized constitutively through the endogenous *Sp. pCas9* promoter [1]. CRISPRi was achieved using *sgRNAs*, expressed from the Anderson *E. coli* promoter J23119. The target template for CRISPRi was the plasmid P70a-*degfp* [2,3]. The reporter protein deGFP, described earlier [2,3], is a slightly truncated version of eGFP with the same fluorescent properties as eGFP. For CRISPR activation experiments in TXTL, we used the same plasmids as for CRISPRi to express *spcas9* or *spdcas9*. The *scRNAs* gene was expressed from the Anderson *E. coli* promoter J23119. The *activator* gene was expressed constitutively from the Anderson *E. coli* promoter J23107 [4,5]. The target template for CRISPRa was the plasmids pJF143.J2, pJF143.J3, and pJF143.J4 that contain the Anderson promoter J23117 [4,5] upstream of the *mrfp* gene. The *sgRNA* and *scRNAs* were expressed using linear templates, whose degradation in TXTL was prevented by adding the Chi6 dsDNA linear template [6]. All the constructs have been sequenced. These plasmids are available on demand.

**TXTL reactions.** The myTXTL kit (Arbor Biosciences) was used for cell-free expression. TXTL reactions are composed of an *E. coli* lysate, an amino acid mixture, an energy buffer, and the desired DNA templates. This TXTL system has been described previously [2,3]. All the batch mode TXTL reactions were incubated at 29 °C in either a bench-top incubator, for endpoint measurements, or in the plate readers, for kinetic measurements. 29 °C is the optimum temperature of incubation for the myTXTL kit.

**Quantitative measurements of fluorescence in TXTL reactions.** Fluorescence from batch mode TXTL reactions was measured using the reporter protein deGFP (25.4 kDa, 1 mg/ml = 39.4 μM) and mRFP (25.4 kDa, 1 mg/ml = 39.4 μM). Fluorescence was measured at five-minute intervals using monochromators (deGFP Ex/Em 488/525 nm, mRFP Ex/Em 555/583) on Biotek Synergy H1 plate readers in Costar polypropylene 96-well, V-bottom plates. Endpoint reactions were measured after 12 h of incubation. To measure protein concentration (eGFP reporter), a linear calibration curve of fluorescence intensity versus eGFP concentration was generated using purified recombinant eGFP obtained from Cell Biolabs (STA-201), Inc or purified in the lab [3]. The fraction of silenced reporters *R* for CRISPRi is calculated as $R = 1 - I_{On\ Target}/I_{Off\ Target}$, where $I_{On\ Target}$ is the intensity of an on target sgRNA TXTL reaction, while $I_{Off\ Target}$ is the intensity of the off target sgRNA TXTL reaction.

**Computer codes.** Scripts are available on the lab Github profile noireauxlab.

**Data availability statement:** Data is available on the lab Github profile noireauxlab.

**Landscape**

In the Michaelis-Menten model used for truncations, the binding is described by a factor $\alpha$, which contains all the parameters unrelated to the binding energy of the guides to the target DNA.

$$p_b = \left(1 + \alpha exp\left(-\beta E_{binding}\right)\right)^{-1}, \alpha = K_{d0}/[CRISPR] \qquad (1)$$

By varying $\alpha$, we can observe different behavior of the same guide for both spdCas9 and spCas9 CRISPRa binding **(Fig. S1a and S1b)**. When the value of $\alpha$ is varied we see a change in the observed dependence of binding with the length of the guide. In the limits of both large ($\alpha$ = 10) and small ($\alpha$ = 10000000) doses of CRISPR, the effects caused by conformational changes become less noticeable.

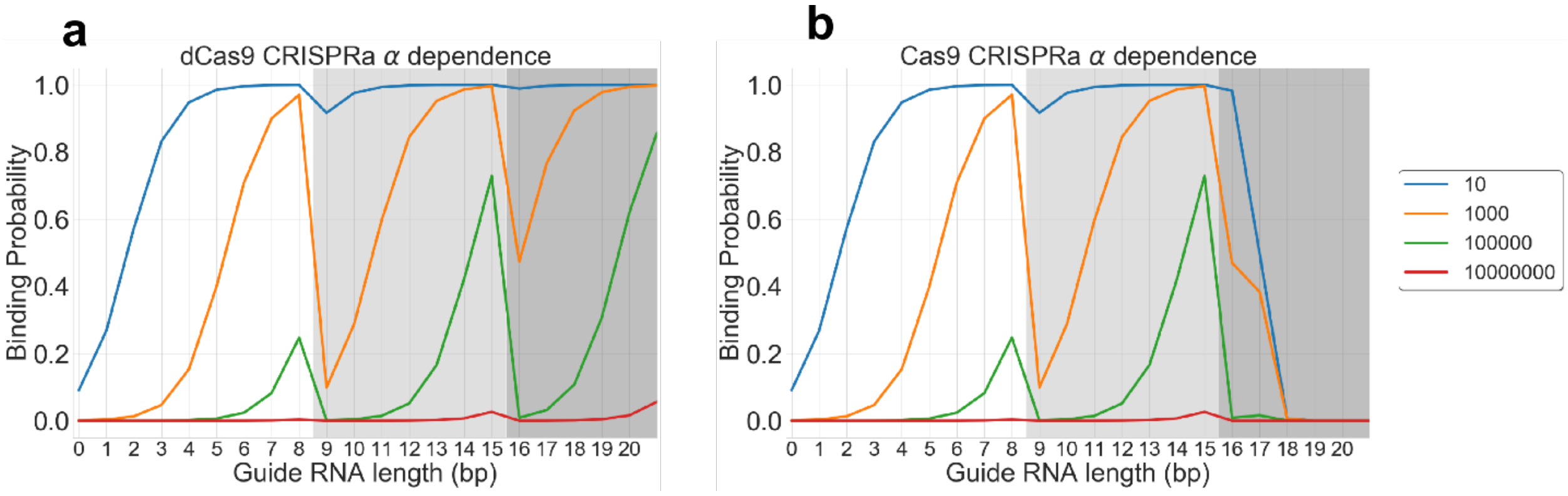

**Fig S1**. CRISPR energy landscape in TXTL. **(a)** and **(b),** the dependence of binding probability on the dimensionless factor $\alpha$ in spdCas9 and spCas9 CRISPRa experiments respectively, where $\alpha$ ranges between 10-1000000.

We performed the CRISPRa experiment with different guide lengths and for different target (promoter) sequences. The scRNA were transcribed from linear DNA templates at different concentrations **(Fig. S2a and S2b)**. We observe that depending on the concentrations of the scRNA template used in the reaction, the peaks caused by pre-conformational stability become more visible. Similarly to CRISPRa, we observe peaks for CRISPRi experiments depending on the sgRNA length and template concentration added to the reaction for another target site, sg9, that targets into the *degfp* gene **(Fig. S2c)**. This indicates that multiple peaks are seen across all targets, but with different characteristics depending on the target sequence. The conformational change concentration dependence was also observed in spCas9-CRISPRa experiments **(Fig. S2d and S2e)**.

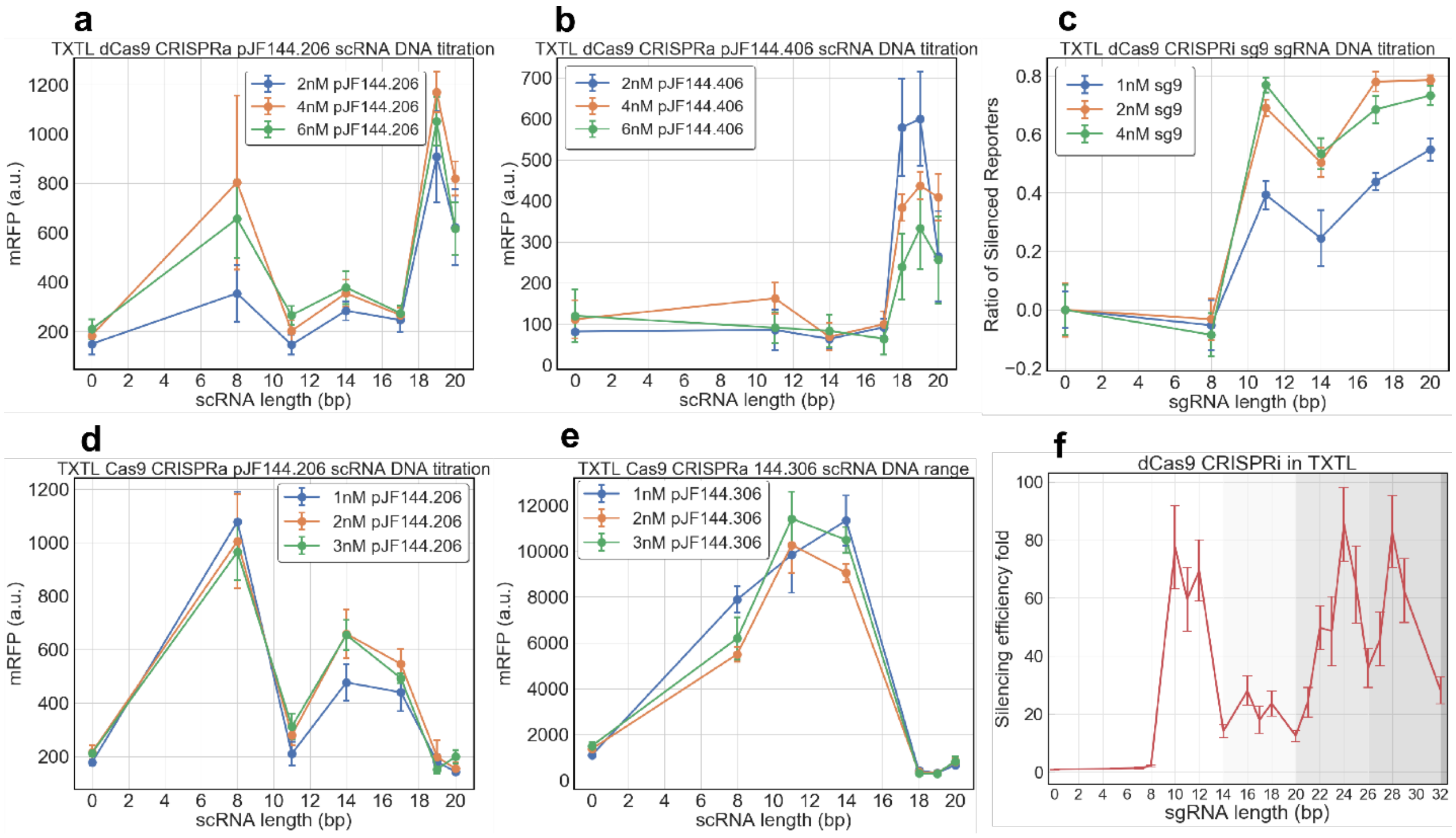

**Fig S2**. Length experiments with other target sequences. **(a)** and **(b),** spdCas9-CRISPRa scRNA length dependence in TXTL for different concentrations of linear scRNA transcribing DNA pJF144.206 and pJF144.406 respectively. **(c),** spdCas9 CRISPRi experiment, sgRNA length dependence in TXTL for different concentrations of linear sgRNA transcribing DNA targeting the *egfp* gene (sg9 in [7]). **(d)** and **(e),** spCas9-CRISPRa scRNA length dependence in TXTL for different concentrations of linear scRNA DNA pJF144.206 and pJF144.306 respectively. **(f),** the fold repression for the spdCas9 CRISPRi experiment **(Fig. 2a)**.

## Mismatch analysis from other studies

Besides single mismatches in spCas9 and Cas12, the model can also be applied to multiple mismatches. Using the energies acquired from single mismatch data for LbCpf1 experiments we constructed a tandem mismatch plot and compared it to tandem-mismatch experiments **(Fig. S3a)** [8]. We observe that the results from single mismatch experiments offer some predictability to multiple mismatches.

The model can also be applied to Cas3 enzymes **(Fig. S3b)** [9]. Cas3 utilizes cascades of enzymes that perform in concert to cut targets. One of the enzymes in the cascade of Cas3 are helicase enzymes, therefore unlike Cas9 and Cas12 CRISPR systems, unwinding is driven by ATP hydrolysis. As expected for an ATP driven process, the unwinding energies are lower than for thermally driven unwindings of Cas9 and Cas12 systems. The original data was expressed in the energetic penalty of each mismatch. We converted the energetic costs of mismatches to relative cleavage activity (RCA) by taking the exponential of the energetic cost:

$$RCA_i = exp(-E_i/k_bT) \tag{2}$$

The best fit was achieved with conformational changes after the 7th, 13th, 19th, 25th, and 31nd base pairs. This demonstrates that the helicase unwinds the target in 6 bp steps. However, between the large binding peaks there are also smaller peaks that are not captured by the model, which might be caused by either unwinding steps of 3 bp or by dynamical emergent effects not observed in our model.

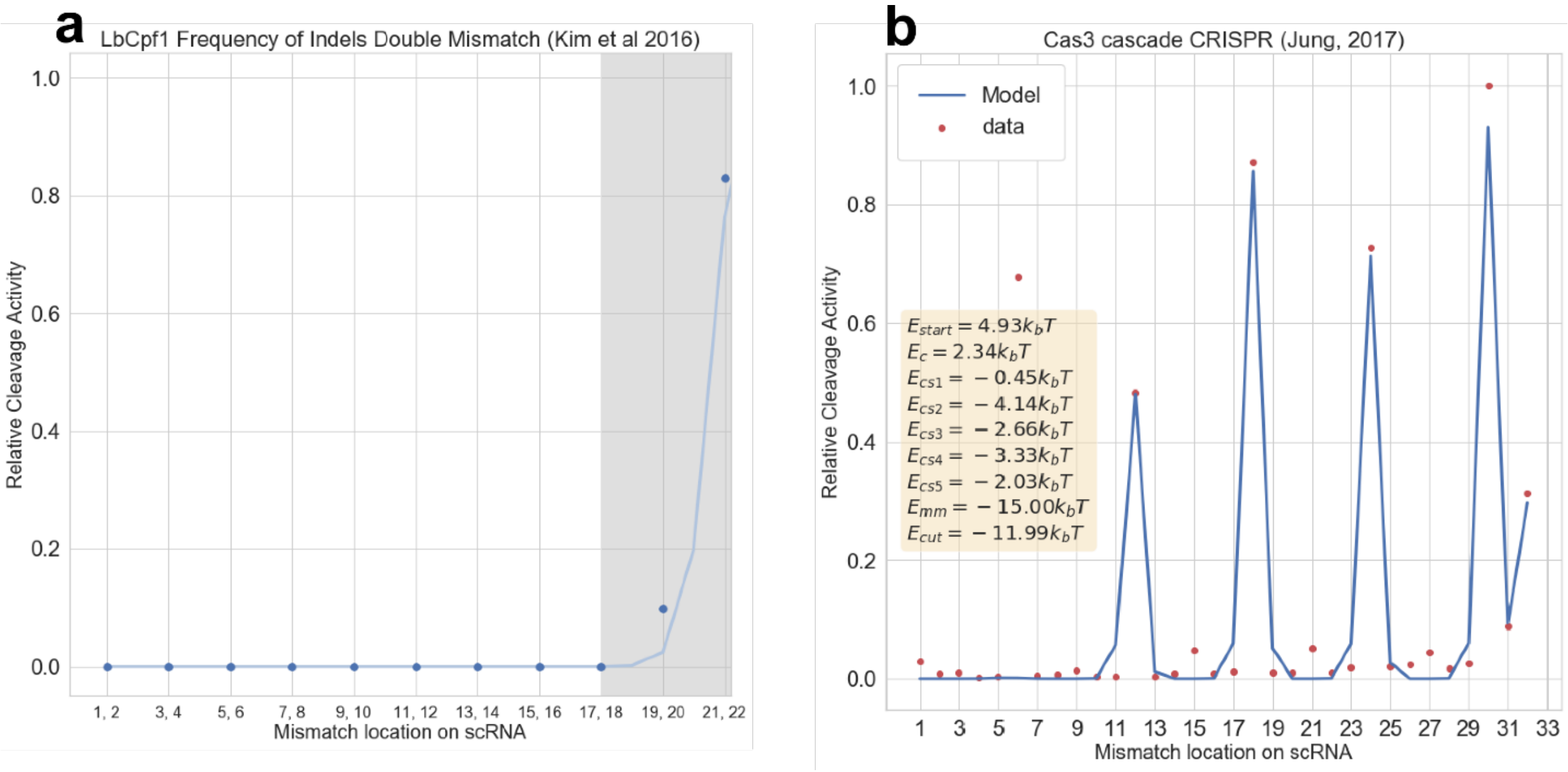

**Fig S3**. Additional mismatch analysis results from published articles. **(a),** applying the mismatch energies acquired from single mismatch results to tandem mismatch data of the same target. **(b),** mismatch analysis in Cas3 Cascade CRISPR system.

Cas13b targets RNA instead of DNA and therefore unwinding the target is no longer necessary since RNA-RNA hybridization will occur simply by allowed proximity. Instead, Cas13b systems vary the availability of guide RNA to the target RNA by changing their conformations, therefore we can still apply the same rules **(Fig. S4a and S4b)** [10]. However, when Cas13b enzymes reach their active conformational state, the active Cas13b start cutting all RNA non-specifically.

Therefore, the analysis implemented for spCas9 and Cas12a does not fully apply to Cas13b and should only offer a qualitative explanation.

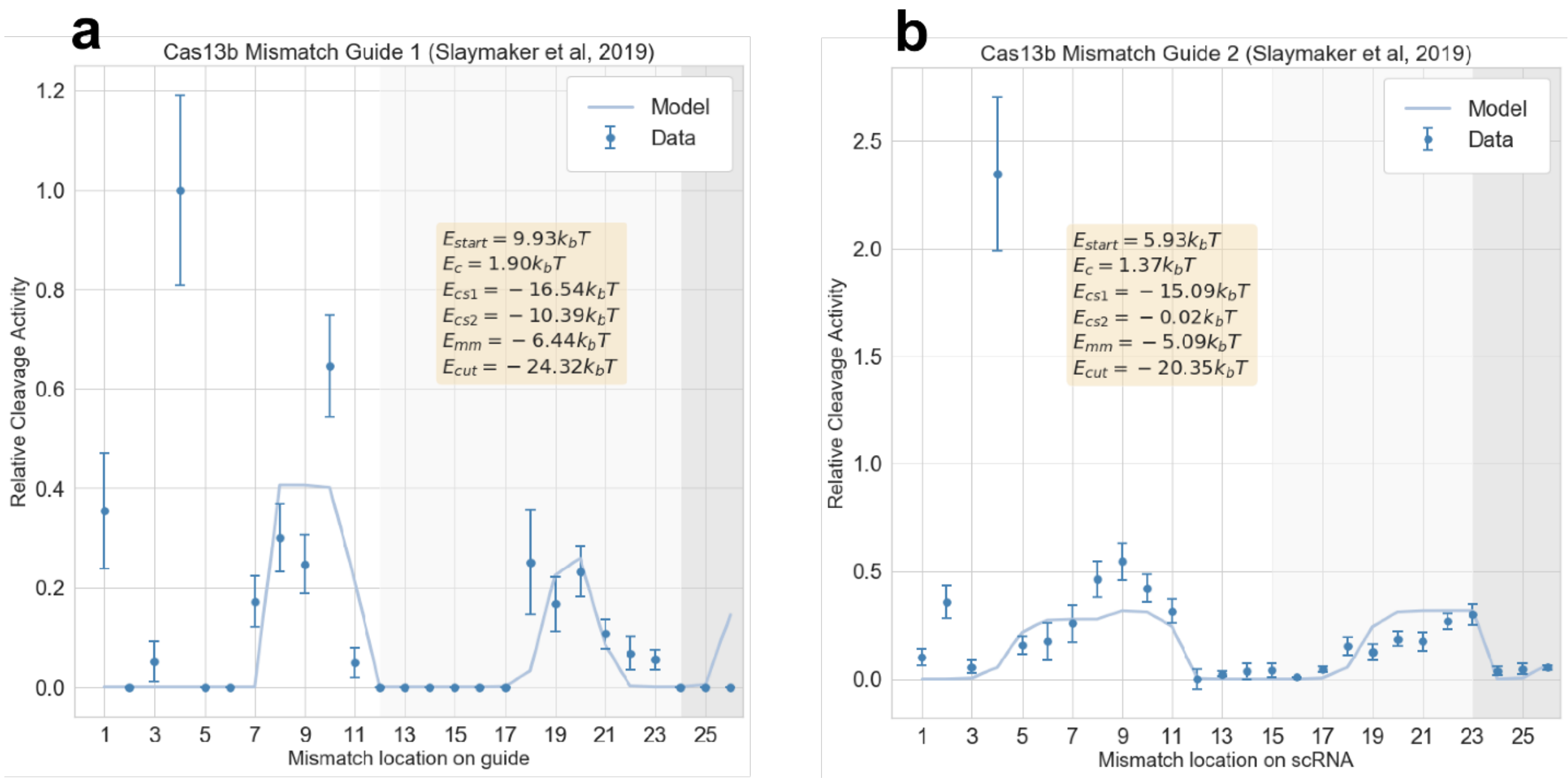

**Fig S4**. Additional mismatch analysis results from published articles. **(a)** and **(b)** mismatch analysis of Cas13b CRISPR systems. Each background shade corresponds to a different conformation of Cas13b that are separated by energetic checkpoints.

**DNA**

The promoters are highlighted in red.
The reporter genes are highlighted in green.
The genes are highlighted in blue.
CRISPR target sequences are highlighted in orange.
The transcription terminator is highlighted in grey.

p70a-deGFP (CRISPRi Reporter)

TGGGCATGCTGAGCTAACACCGTGCGTGTTGACAATTTTACCTCTGGCGGTGATAATGGTT
GCAGCTAGCAATAATTTTGTTTAACTTTAAGAAGGAGATATACCATGGAGCTTTTCACTGGC
GTTGTTCCCATCCTGGTCGAGCTGGACGGCGACGTAAACGGCCACAAGTTCAGCGTGTCC
GGCGAGGGCGAGGGCGATGCCACCTACGGCAAGCTGACCCTGAAGTTCATCTGCACCAC
CGGCAAGCTGCCCGTGCCCTGGCCCACCCTCGTGACCACCCTGACCTACGGCGTGCAGT
GCTTCAGCCGCTACCCCGACCACATGAAGCAGCACGACTTCTTCAAGTCCGCCATGCCCG
AAGGCTACGTCCAGGAGCGCACCATCTTCTTCAAGGACGACGGCAACTACAAGACCCGCG
CCGAGGTGAAGTTCGAGGGCGACACCCTGGTGAACCGCATCGAGCTGAAGGGCATCGAC
TTCAAGGAGGACGGCAACATCCTGGGGCACAAGCTGGAGTACAACTACAACAGCCACAAC
GTCTATATCATGGCCGACAAGCAGAAGAACGGCATCAAGGTGAACTTCAAGATCCGCCAC
AACATCGAGGACGGCAGCGTGCAGCTCGCCGACCACTACCAGCAGAACACCCCCATCGG
CGACGGCCCCGTGCTGCTGCCCGACAACCACTACCTGAGCACCCAGTCCGCCCTGAGCA
AAGACCCCAACGAGAAGCGCGATCACATGGTCCTGCTGGAGTTCGTGACCGCCGCCGGG
ATCTAACTCGAGCAAAGCCCGCCGAAAGGCGGGCTTTTCTGTGTCGA

pJF143.J2 (CRISPRa Reporter 1)

CTGTCGCGCGACTGACCTGACGCATCCTGAGGACGTGTTCGGCTACTACACAAGTATTAA
GAGACAATGCGCTCttgacagctagctcagtcctagggattgtgctagcGAATTCATTAAAGAGGAGAAAGG
TACCATGGCGAGTAGCGAAGACGTTATCAAAGAGTTCATgcgtttcaaagttcgtatggaaggttccgttaac
ggtcacgagttcgaaatcgaaggtgaaggtgaaggtcgtccgtacgaaggtacccagaccgctaaactgaaagttaccaaaggtg
gtccgctgccgttcgcttgggacatcctgtccccgcagttccagtacggttccaaagcttacgttaaacacccggctgacatcccggact
acctgaaactgtccttcccggaaggtttcaaatgggaacgtgttatgaacttcgaagacggtggtgttgttaccgttacccaggactcctc
cctgcaagacggtgagttcatctacaaagttaaactgcgtggtaccaacttcccgtccgacggtccggttatgcagaaaaaaaccatg
ggttgggaagcttccaccgaacgtatgtacccggaagacggtgctctgaaaggtgaaatcaaaatgcgtctgaaactgaaagacgg
tggtcactacgacgctgaagttaaaaccacctacatggctaaaaaaccggttcagctgccgggtgcttacaaaaccgacatcaaact
ggacatcacctcccacaacgaagactacaccatcgttgaacagtacgaacgtgctgaaggtcgtcactccaccggtgcttaaggatc
caaactcgagtaaggatctCCAGGCATCAAATAAAACGAAAGGCTCAGTCGAAAGACTGGGCCTTT
CGTTTTATCTGTTGTTTGTCGGTGAACGCTCTCTACTAGAGTCACACTGGCTCACCTTCGG
GTGGGCCTTTCTGCGTTTATACct

pJF143.J3 (CRISPRa Reporter 2)

CGCGTTCGCTCGTCTCCTCACTTCTCCTACGGAGCGTTCTGGACACAACGTCGTCTTGAAG
TTGCGATTATAGAttgacagctagctcagtcctagggattgtgctagcGAATTCATTAAAGAGGAGAAAGGTA
CCATGGCGAGTAGCGAAGACGTTATCAAAGAGTTCATgcgtttcaaagttcgtatggaaggttccgttaacgg
tcacgagttcgaaatcgaaggtgaaggtgaaggtcgtccgtacgaaggtacccagaccgctaaactgaaagttaccaaaggtggtc
cgctgccgttcgcttgggacatcctgtccccgcagttccagtacggttccaaagcttacgttaaacacccggctgacatcccggactac
ctgaaactgtccttcccggaaggtttcaaatgggaacgtgttatgaacttcgaagacggtggtgttgttaccgttacccaggactcctccc
tgcaagacggtgagttcatctacaaagttaaactgcgtggtaccaacttcccgtccgacggtccggttatgcagaaaaaaaccatggg
ttgggaagcttccaccgaacgtatgtacccggaagacggtgctctgaaaggtgaaatcaaaatgcgtctgaaactgaaagacggtg
gtcactacgacgctgaagttaaaaccacctacatggctaaaaaaccggttcagctgccgggtgcttacaaaaccgacatcaaactg
gacatcacctcccacaacgaagactacaccatcgttgaacagtacgaacgtgctgaaggtcgtcactccaccggtgcttaaggatcc
aaactcgagtaaggatctCCAGGCATCAAATAAAACGAAAGGCTCAGTCGAAAGACTGGGCCTTTC

GTTTTATCTGTTGTTTGTCGGTGAACGCTCTCTACTAGAGTCACACTGGCTCACCTTCGGG
TGGGCCTTTCTGCGTTTATACct

pCD017_dCas9_Endo_promoter

ttaagTTACGAAATCATCCTGTGGAGCTTAGTAGGTTTAGCAAGATGGCAGCGCCTAAATGTA
GAATGATAAAAGGATTAAGAGATTAATTTCCCTAAAAATGATAAAACAAGCGTTTTGAAAGC
GCTTGTTTTTTTGGTTTGCAGTCAGAGTAGAATAGAAGTATCAAAAAAAGCACCGACTCGGT
GCCACTTTTTCAAGTTGATAACGGACTAGCCTTATTTTAACTTGCTATGCTGTTTTGAATGGT
TCCAACAAGATTATTTTATAACTTTTATAACAAATAATCAAGGAGAAATTCAAAGAAATTTATC
AGCCATAAAACAATACTTAATACTATAGAATGATAACAAAATAAACTACTTTTTAAAAGAATTT
TGTGTTATAATCTATTTATTATTAAGTATTGGGTAATATTTTTTGAAGAGATATTTTGAAAAAG
AAAAATTAAAGCATATTAAACTAATTTCGGAGGTCATTAAAACTATTATTGAAATCATCAAAC
TCATTATGGATTTAATTTAAACTTTTTATTTTAGGAGGCAAAAATGGATAAGAAATACTCAAT
AGGCTTAGcTATCGGCACAAATAGCGTCGGATGGGCGGTGATCACTGATGAATATAAGGTT
CCGTCTAAAAAGTTCAAGGTTCTGGGAAATACAGACCGCCACAGTATCAAAAAAAATCTTAT
AGGGGCTCTTTTATTTGACAGTGGAGAGACAGCGGAAGCGACTCGTCTCAAACGGACAGC
TCGTAGAAGGTATACACGTCGGAAGAATCGTATTTGTTATCTACAGGAGATTTTTTCAAATG
AGATGGCGAAAGTAGATGATAGTTTCTTTCATCGACTTGAAGAGTCTTTTTTGGTGGAAGAA
GACAAGAAGCATGAACGTCATCCTATTTTTGGAAATATAGTAGATGAAGTTGCTTATCATGA
GAAATATCCAACTATCTATCATCTGCGAAAAAAATTGGTAGATTCTACTGATAAAGCGGATT
TGCGCTTAATCTATTTGGCCTTAGCGCATATGATTAAGTTTCGTGGTCATTTTTTGATTGAG
GGAGATTTAAATCCTGATAATAGTGATGTGGACAAACTATTTATCCAGTTGGTACAAACCTA
CAATCAATTATTTGAAGAAAACCCTATTAACGCAAGTGGAGTAGATGCTAAAGCGATTCTTT
CTGCACGATTGAGTAAATCAAGACGATTAGAAAATCTCATTGCTCAGCTCCCCGGTGAGAA
GAAAAATGGCTTATTTGGGAATCTCATTGCTTTGTCATTGGGTTTGACCCCTAATTTTAAATC
AAATTTTGATTTGGCAGAAGATGCTAAATTACAGCTTTCAAAAGATACTTACGATGATGATTT
AGATAATTTATTGGCGCAAATTGGAGATCAATATGCTGATTTGTTTTTGGCAGCTAAGAATT
TATCAGATGCTATTTTACTTTCAGATATCCTAAGAGTAAATACTGAAATAACTAAGGCTCCCC
TATCAGCTTCAATGATTAAACGCTACGATGAACATCATCAAGACTTGACTCTTTTAAAAGCTT
TAGTTCGACAACAACTTCCAGAAAAGTATAAAGAAATCTTTTTTGATCAATCAAAAAACGGAT
ATGCAGGTTATATTGATGGGGGAGCTAGCCAAGAAGAATTTTATAAATTTATCAAACCAATT
TTAGAAAAAATGGATGGTACTGAGGAATTATTGGTGAAACTAAATCGTGAAGATTTGCTGCG
CAAGCAACGGACCTTTGACAACGGCTCTATTCCCCATCAAATTCACTTGGGTGAGCTGCAT
GCTATTTTGAGAAGACAAGAAGACTTTTATCCATTTTTAAAAGACAATCGTGAGAAGATTGA
AAAAATCTTGACTTTTCGAATTCCTTATTATGTTGGTCCATTGGCGCGTGGCAATAGTCGTT
TTGCATGGATGACTCGGAAGTCTGAAGAAACAATTACCCCATGGAATTTTGAAGAAGTTGT
CGATAAAGGTGCTTCAGCTCAATCATTTATTGAACGCATGACAAACTTTGATAAAAATCTTC
CAAATGAAAAAGTACTACCAAAACATAGTTTGCTTTATGAGTATTTTACGGTTTATAACGAAT
TGACAAAGGTCAAATATGTTACTGAAGGAATGCGAAAACCAGCATTTCTTTCAGGTGAACA
GAAGAAAGCCATTGTTGATTTACTCTTCAAAACAAATCGAAAAGTAACCGTTAAGCAATTAA
AAGAAGATTATTTCAAAAAAATAGAATGTTTTGATAGTGTTGAAATTTCAGGAGTTGAAGATA
GATTTAATGCTTCATTAGGTACCTACCATGATTTGCTAAAAATTATTAAAGATAAAGATTTTTT
GGATAATGAAGAAAATGAAGATATCTTAGAGGATATTGTTTTAACATTGACCTTATTTGAAGA
TAGGGAGATGATTGAGGAAAGACTTAAAACATATGCTCACCTCTTTGATGATAAGGTGATG
AAACAGCTTAAACGTCGCCGTTATACTGGTTGGGGACGTTTGTCTCGAAAATTGATTAATG
GTATTAGGGATAAGCAATCTGGCAAAACAATATTAGATTTTTTGAAATCAGATGGTTTTGCC
AATCGCAATTTTATGCAGCTGATCCATGATGATAGTTTGACATTTAAAGAAGACATTCAAAA
AGCACAAGTGTCTGGACAAGGCGATAGTTTACATGAACATATTGCAAATTTAGCTGGTAGC
CCTGCTATTAAAAAAGGTATTTTACAGACTGTAAAAGTTGTTGATGAATTGGTCAAAGTAAT
GGGGCGGCATAAGCCAGAAAATATCGTTATTGAAATGGCACGTGAAAATCAGACAACTCAA
AAGGGCCAGAAAAATTCGCGAGAGCGTATGAAACGAATCGAAGAAGGTATCAAAGAATTAG

GAAGTCAGATTCTTAAAGAGCATCCTGTTGAAAATACTCAATTGCAAAATGAAAAGCTCTAT
CTCTATTATCTCCAAAATGGAAGAGACATGTATGTGGACCAAGAATTAGATATTAATCGTTT
AAGTGATTATGATGTCGATgcCATTGTTCCACAAAGTTTCCTTAAAGACGATTCAATAGACAA
TAAGGTCTTAACGCGTTCTGATAAAAATCGTGGTAAATCGGATAACGTTCCAAGTGAAGAA
GTAGTCAAAAAGATGAAAAACTATTGGAGACAACTTCTAAACGCCAAGTTAATCACTCAACG
TAAGTTTGATAATTTAACGAAAGCTGAACGTGGAGGTTTGAGTGAACTTGATAAAGCTGGTT
TTATCAAACGCCAATTGGTTGAAACTCGCCAAATCACTAAGCATGTGGCACAAATTTTGGAT
AGTCGCATGAATACTAAATACGATGAAAATGATAAACTTATTCGAGAGGTTAAAGTGATTAC
CTTAAAATCTAAATTAGTTTCTGACTTCCGAAAAGATTTCCAATTCTATAAAGTACGTGAGAT
TAACAATTACCATCATGCCCATGATGCGTATCTAAATGCCGTCGTTGGAACTGCTTTGATTA
AGAAATATCCAAAACTTGAATCGGAGTTTGTCTATGGTGATTATAAAGTTTATGATGTTCGTA
AAATGATTGCTAAGTCTGAGCAAGAAATAGGCAAAGCAACCGCAAAATATTTCTTTTACTCT
AATATCATGAACTTCTTCAAAACAGAAATTACACTTGCAAATGGAGAGATTCGCAAACGCCC
TCTAATCGAAACTAATGGGGAAACTGGAGAAATTGTCTGGGATAAAGGGCGAGATTTTGCC
ACAGTGCGCAAAGTATTGTCCATGCCCCAAGTCAATATTGTCAAGAAAACAGAAGTACAGA
CAGGCGGATTCTCCAAGGAGTCAATTTTACCAAAAAGAAATTCGGACAAGCTTATTGCTCG
TAAAAAAGACTGGGATCCAAAAAAATATGGTGGTTTTGATAGTCCAACGGTAGCTTATTCAG
TCCTAGTGGTTGCTAAGGTGGAAAAAGGGAAATCGAAGAAGTTAAAATCCGTTAAAGAGTT
ACTAGGGATCACAATTATGGAAAGAAGTTCCTTTGAAAAAAATCCGATTGACTTTTTAGAAG
CTAAAGGATATAAGGAAGTTAAAAAAGACTTAATCATTAAACTACCTAAATATAGTCTTTTTG
AGTTAGAAAACGGTCGTAAACGGATGCTGGCTAGTGCCGGAGAATTACAAAAAGGAAATGA
GCTGGCTCTGCCAAGCAAATATGTGAATTTTTTATATTTAGCTAGTCATTATGAAAAGTTGAA
GGGTAGTCCAGAAGATAACGAACAAAAACAATTGTTTGTGGAGCAGCATAAGCATTATTTA
GATGAGATTATTGAGCAAATCAGTGAATTTTCTAAGCGTGTTATTTTAGCAGATGCCAATTT
AGATAAAGTTCTTAGTGCATATAACAAACATAGAGACAAACCAATACGTGAACAAGCAGAAA
ATATTATTCATTTATTTACGTTGACGAATCTTGGAGCTCCCGCTGCTTTTAAATATTTTGATA
CAACAATTGATCGTAAACGATATACGTCTACAAAAGAAGTTTTAGATGCCACTCTTATCCAT
CAATCCATCACTGGTCTTTATGAAACACGCATTGATTTGAGTCAGCTAGGAGGTGACTAACt
cgagtaaggatctccaggcatcaaataaaacgaaaggctcagtcgaaagactgggcctttcgttttatctgttgtttgtcggtgaacgct
ctctactagagtcacactggctcaccttcgggtgggcctttctgcgtttataacctag

pBT005:MCP_Sox (Activator)

ctgcagtttacggctagctcagccctaggtattatgctagcGAATTCATTAAAGAGGAGAAAGGTACCatggggccc
gcttctaactttactcagttcgttctcgtcgacaatggcggaactggcgacgtgactgtcgccccaagcaacttcgctaacgggatcgct
gaatggatcagctctaactcgcgttcacaggcttacaaagtaacctgtagcgttcgtcagagctctgcgcagaatcgcaaatacacca
tcaaagtcgaggtgcctaaaggcgcctggcgttcgtacttaaatatggaactaaccattccaattttcgccacgaattccgactgcgag
cttattgttaaggcaatgcaaggtctcctaaaagatggaaacccgattccctcagcaatcgcagcaaactccggcatctacGGTGG
CGGAGGTAGCATGTCCCATCAGAAAATTATTCAGGATCTTATCGCATGGATTGACGAGCAT
ATTGACCAGCCGCTTAACATTGATGTAGTCGCAAAAAAATCAGGCTATTCAAAGTGGTACTT
GCAACGAATGTTCCGCACGGTGACGCATCAGACGCTTGGCGATTACATTCGCCAACGCCG
CCTGTTACTGGCCGCCGTTGAGTTGCGCACCACCGAGCGTCCGATTTTTGATATCGCAATG
GACCTGGGTTATGTCTCGCAGCAGACCTTCTCCCGCGTTTTCGCGCGGCAGTTTGATCGC
ACTCCCAGCGATTATCGCCACCGCCTGTAAGCGGCCGCcacgcaaaaaccccgcttcggcggggtttttt
cgc

pCas9

https://www.addgene.org/42876/

Spy-sgRNA-anti-P70a-pos6-20

CGAGAGCATCGATCCTAGCATGCGTCTCATCGGGatgcatTTGACAGCTAGCTCAGTCCTAG
GTATAATGCTAGCGGTAAAATTGTCAACACGCAGTTTTAGAGCTAGAAATAGCAAGTTAAAA
TAAGGCTAGTCCGTTATCAACTTGAAAAAGTGGCACCGAGTCGGTGCTTTTTTTccggcttatcg
gtcagtttcacctgatttacgtaaaaacccgcttcggcgggtttttgcttttggaggggcagaaagatgaatgactgtccacgacgctata
cccaaaagaaaaagcttGAAAGAAAGAATCGATCGATCCCGAATGAGTAG

Spy-sgRNA-anti-P70a-pos9-20

CGAGAGCATCGATCCTAGCATGCGTCTCATCGGGatgcatTTGACAGCTAGCTCAGTCCTAG
GTATAATGCTAGCGTCGCCCTCGAACTTCACCTGTTTTAGAGCTAGAAATAGCAAGTTAAAA
TAAGGCTAGTCCGTTATCAACTTGAAAAAGTGGCACCGAGTCGGTGCTTTTTTTccggcttatcg
gtcagtttcacctgatttacgtaaaaacccgcttcggcgggtttttgcttttggaggggcagaaagatgaatgactgtccacgacgctata
cccaaaagaaaaagcttGAAAGAAAGAATCGATCGATCCCGAATGAGTAG

144.306.20 scRNA

tttctagatttcagtgcaatttatctcttcaaatgtagcacctgaagtcagccccatacgatATAAGTTGTTACTAGATTGAC
AGCTAGCTCAGTCCTAGGTATAATACTAGTTTGTGTCCAGAACGCTCCGTGTTTTAGAGCTA
GAAATAGCAAGTTAAAATAAGGCTAGTCCGTTATCAACTTGAAAAAGTGGCACATGAGGAT
CACCCATGTGCTTTTTTTGAAGCTTGGGCCCGAACAAAAACTCATCTCAGAAGAGGATCTG
AATAGCGCCGTCGACCATCATCATCATCATCATTGAGTTTAAACGGTCTCCAGCTTGGCTG
TTTTGGCGGATGAGAGAAGATTTTCAGCCTGATACAGATTAAATCAGAACGCAGAAGCGGT
CTGATAAAACAGAATTTGCCTGGCGGCAGTAGCGCGGTGGTCCCACCTGACCCCATGCCG
AACTCAGAAGTGAAACGCCGTAGCGCCGATGGTAGTGTGGGGTCTCCCCATGCGAGAGTA
GGGAACTGCCAGGCATCAAATAAAACGAAAGGCTCAGTCGAAAGACTGGGCCTTTCGTTTT
ATCTGTTGTTTGTCGGTGAACTGGATCCTTACTCGAGTCTAGACT

144.206.20 scRNA

tttctagatttcagtgcaatttatctcttcaaatgtagcacctgaagtcagccccatacgatATAAGTTGTTACTAGATTGAC
AGCTAGCTCAGTCCTAGGTATAATACTAGTTAGTAGCCGAACACGTCCTCGTTTTAGAGCT
AGAAATAGCAAGTTAAAATAAGGCTAGTCCGTTATCAACTTGAAAAAGTGGCACATGAGGA
TCACCCATGTGCTTTTTTTGAAGCTTGGGCCCGAACAAAAACTCATCTCAGAAGAGGATCT
GAATAGCGCCGTCGACCATCATCATCATCATCATTGAGTTTAAACGGTCTCCAGCTTGGCT
GTTTTGGCGGATGAGAGAAGATTTTCAGCCTGATACAGATTAAATCAGAACGCAGAAGCGG
TCTGATAAAACAGAATTTGCCTGGCGGCAGTAGCGCGGTGGTCCCACCTGACCCCATGCC
GAACTCAGAAGTGAAACGCCGTAGCGCCGATGGTAGTGTGGGGTCTCCCCATGCGAGAGT
AGGGAACTGCCAGGCATCAAATAAAACGAAAGGCTCAGTCGAAAGACTGGGCCTTTCGTTT
TATCTGTTGTTTGTCGGTGAACTGGATCCTTACTCGAGTCTAGACT